\documentclass{PoS}

\usepackage{aas_macros}
\title{Limit on an Isotropic Diffuse Gamma-Ray Population with HAWC}

\ShortTitle{Isotropic Diffuse Gamma-Ray Emission in HAWC}

\author{\speaker{John Pretz}$^a$ for the HAWC Collaboration$^b$ \\
        \llap{$^a$}Department of Physics, Pennsylvania State University, State College, PA, USA \\
        \llap{$^b$}For a complete author list, see \href{http://www.hawc-observatory.org/collaboration/icrc2015.php}{www.hawc-observatory.org/collaboration/icrc2015.php} \\
        Email: \email{pretz@psu.edu}}

\abstract{Data from 105 days from the High Altitude Water Cherenkov
Observatory (HAWC) have been used to place a new limit on an
isotropic diffuse gamma-ray population above 10 TeV.
High-energy isotropic diffuse gamma-ray emission is produced 
by unresolved extragalactic objects such as active galactic nuclei, 
with potential contributions from interactions of high-energy cosmic rays
with the inter-Galactic medium, or dark matter annihilation. 
Isotropic 
diffuse gamma-ray emission has been observed up to nearly 1 TeV. Above
this energy, only upper limits have been reported. Observations or 
limits of the isotropic photon population above these energies are very
sensitive to local astrophysical particle production. Of particular note,
we expect a photon population to accompany the TeV-PeV astrophysical
neutrino detection seen in the IceCube instrument.
Observations or limits of a photon population 
above this energy can point to the origin of these neutrinos, indicating 
whether they
are within the gamma-ray horizon or not. 
HAWC, with superior 
sensitivity to gamma rays between 100 GeV and 100 TeV, continuously 
observes the overhead sky and will measure or constrain isotropic emission 
above 1 TeV. We present a limit above 10 TeV
based on the background rejection achieved in a 105-day observation
of the Crab Nebula with HAWC. 
The limit will improve substantially with additional
data and study.}

\FullConference{The 34th International Cosmic Ray Conference,\\
		30 July- 6 August, 2015\\
		The Hague, The Netherlands}

\begin{document}

\section{Motivation}
\label{introsection}

Isotropic diffuse gamma-ray emission can arise from a number of mechanisms
including unresolved sources,
dark matter\cite{isodm2001}\cite{isodm2012}\cite{isodm2002},
or more exotic
phenomena\cite{isotheorycrazyshit1992}\cite{isotheorycrazyshit1994}\cite{isotheorycrazyshit1996}.
Of particular interest are gamma-rays produced in
conjunction with the TeV-PeV neutrino population reported by the
IceCube collaboration
\cite{icecubenu2013}\cite{icecubenu2014a}\cite{icecubenu2014b}
which may occur in
astrophysical pion production or
in dark matter annihilation \cite{darkmatterneutrino2015}.
Charged pion
production should be accompanied by neutral
pion production and
these neutral pions will decay into gamma rays.
If the IceCube neutrinos
are from pion decay,
gamma rays should be produced with a similar spectrum and flux
as the neutrinos \cite{neutrinocase1995}\cite{milagrocygnusneutrinos2007}.
Gamma rays above 1 TeV that originate outside the galaxy are heavily attenuated
by the Extra-Galactic Background Light (EBL).
This situation
creates a testable hypothesis: A photon flux at the level of the
IceCube measurement suggest that the neutrinos are produced locally, within
or near our own galaxy \cite{2014PhRvD..90b3010A}. 
An absence of photons would suggest the neutrinos are
produced beyond the gamma-ray horizon \cite{2013PhRvD..88l1301M}. 
For this reason, diffuse
gamma-ray observations are particularly interesting if they reach 
sensitivity below the level
of the IceCube astrophysical neutrino detection.

Electrons and positrons produce similar air
showers to gamma rays and appear very similar in a ground array.
Cosmic-ray electrons can be accelerated by a nearby pulsar or
supernova remnant, but the electrons are isotropized by the intervening
magnetic field.
Electrons
have been observed up to 1 TeV by H.E.S.S. \cite{hesscrelectrons}.
The spectrum appears to be falling
very rapidly, near $E^{-4}$.
The cosmic-ray electron/positron spectrum is interesting in its own
right; models suggest that the cosmic-ray electron spectrum may flatten if a
nearby source is bright enough
\cite{snrelectrons2004}\cite{pulsarelectrons2004}. It may also
constitute a background to photon observation if the spectrum 
hardens.

The High Altitude Water Cherenkov Observatory (HAWC) is a water-Cherenkov
air shower array operating near Sierra Negra in Mexico. A key feature
of the high-energy sensitivity of the HAWC instrument is its
ability to distinguish electromagnetic air showers
from hadronic showers. 
The 
300 
4.5-meter deep 
Water Cherenkov Detectors (WCDs) that make up the instrument 
are designed for identifying the penetrating
particles characteristic of hadronic air showers. HAWC maintains this ability
across its entire 20,000 ${\rm{m}^2}$ area.

HAWC is designed to observe localized sources of TeV photons. When
observing a localized source, the hadron background is measured
from data by interpolating
between
``off source'' observations. When looking for an isotropic population,
the hadron background must be determined {\it{a priori}} from simulation. 
HAWC's ability to distinguish hadronic showers from 
electromagnetic showers is critical. 
%In Section \ref{simplecalculations},
%we make some simple arguments to demonstrate the rough
%order-of-magnitude photon/hadron
%separation required to be able to make meaningful 
%constraints on the isotropic diffuse emission. 
In Section \ref{crabstudies}
we present some studies of the photon/hadron rejection by looking 
at the 10 TeV signal from the Crab Nebula. 
The hadron rejection is so
good that we can improve upon existing isotropic photon 
limits at $\sim$10 TeV 
without a dedicated isotropic photon analysis. In Section \ref{thelimit},
we show the computation of this limit. A number of 
analysis improvements
stand to improve our sensitivity by a factor of 100. Section 
\ref{discussionandprospects} discusses these prospects.

HAWC cannot easily distinguish between high-energy electrons,
positrons and photons.
To the extent that air showers
from electrons, positrons and photons behave the same, our 
limits are actually a limit on the combined flux of photons, electrons
and positrons. 
For this analysis, we frame the discussion in terms of gamma rays
because we have not yet considered any 
differences in the air showers generated by electrons or positrons.
The prospects for detecting the $\sim$TeV electron/positron population
are discussed elsewhere in these proceedings \cite{segevsproceedings}.

\section{Crab Studies}
\label{crabstudies}

As the strongest steady TeV source in the sky, the Crab Nebula serves as an important
benchmark and calibration source for TeV instruments, including HAWC. 
The Crab is detected in HAWC data (in the 250-tank configuration) 
at 38 standard deviations in 150 days of observation
\cite{pacosproceedings} and is being used to verify the pointing, stability
and energy response of the HAWC instrument. We also use the 
Crab to evaluate the effectiveness of our photon/hadron discrimination.

\begin{figure}
\includegraphics[width=0.95\linewidth]{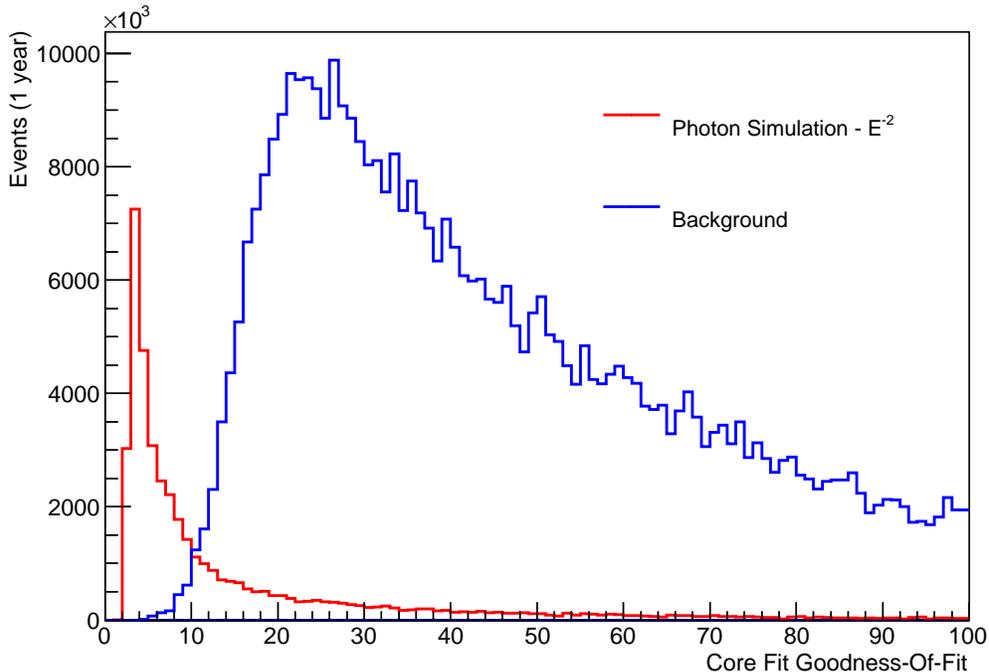}
\caption{Background separation at 10 TeV with HAWC. The number
of events per year is shown
as a function of
the goodness-of-fit for simulated cosmic rays and gamma rays in 
HAWC-250. 
Events with more than 85\% of the available PMTs are used.
The gamma rays are shown
with a simulated spectrum 
$E^{2}\Phi(E)=2000\cdot3\times10^{-8}~{\rm{GeV}}~{\rm{s}}^{-1}~{\rm{cm}}^{-2}~{\rm{sr}}^{-2}$,
2000 times larger than the IceCube neutrino flux. }
\label{corechisq}
\end{figure}

Figure \ref{corechisq} exhibits one photon identification variable,
a goodness-of-fit
measure
of the HAWC core reconstruction, described in \cite{andysproceedings}.
It provides excellent
separation of photons and hadrons, especially at the high energies
we are using for this isotropic photon analysis. The figure corresponds
to a photon energy of more than 10 TeV.
Photons characteristically
have a much lower (i.e. better) goodness-of-fit than cosmic rays, 
allowing us
to distinguish the two.

\begin{figure}
\includegraphics[width=0.48\linewidth]{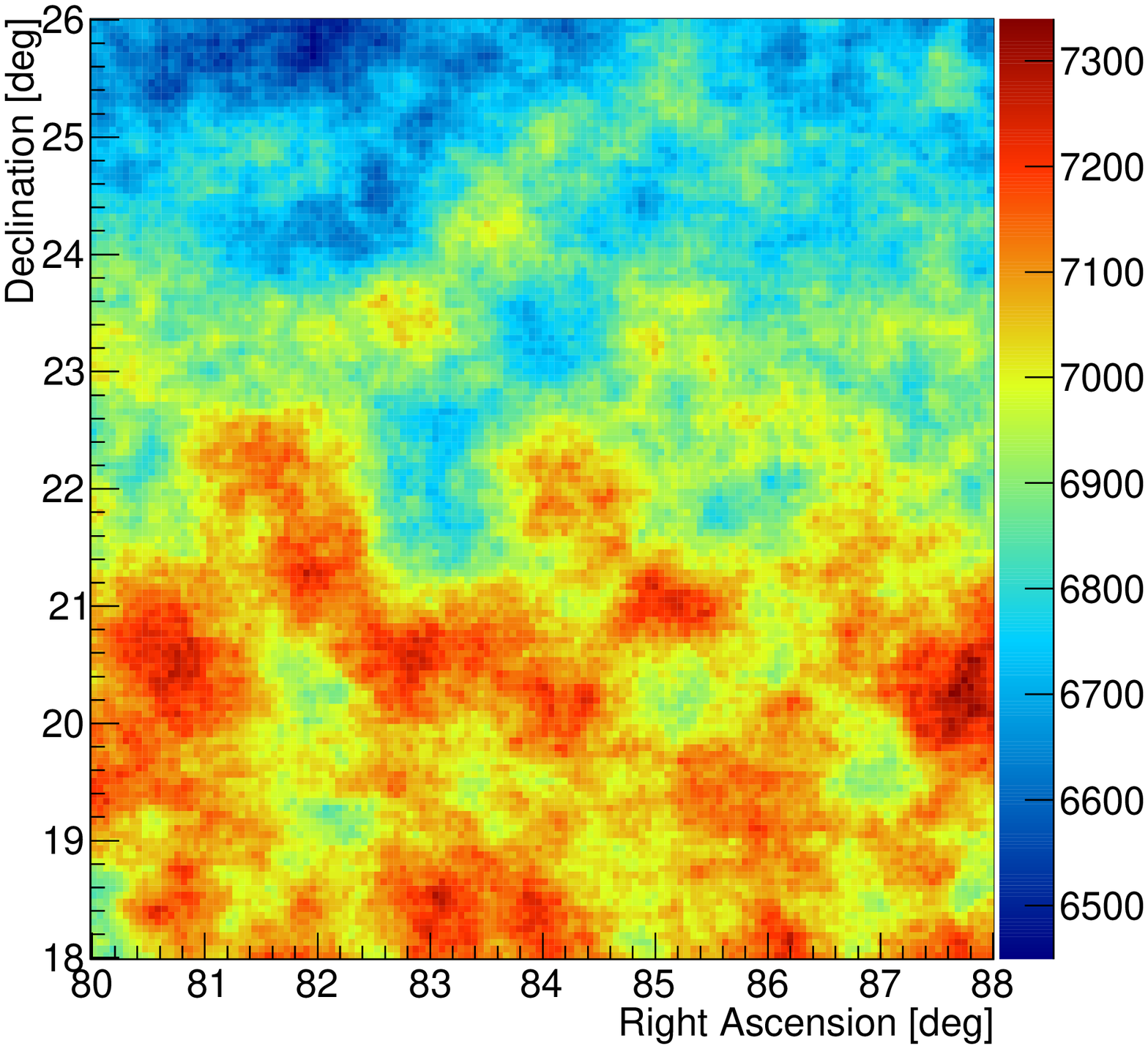}
\includegraphics[width=0.48\linewidth]{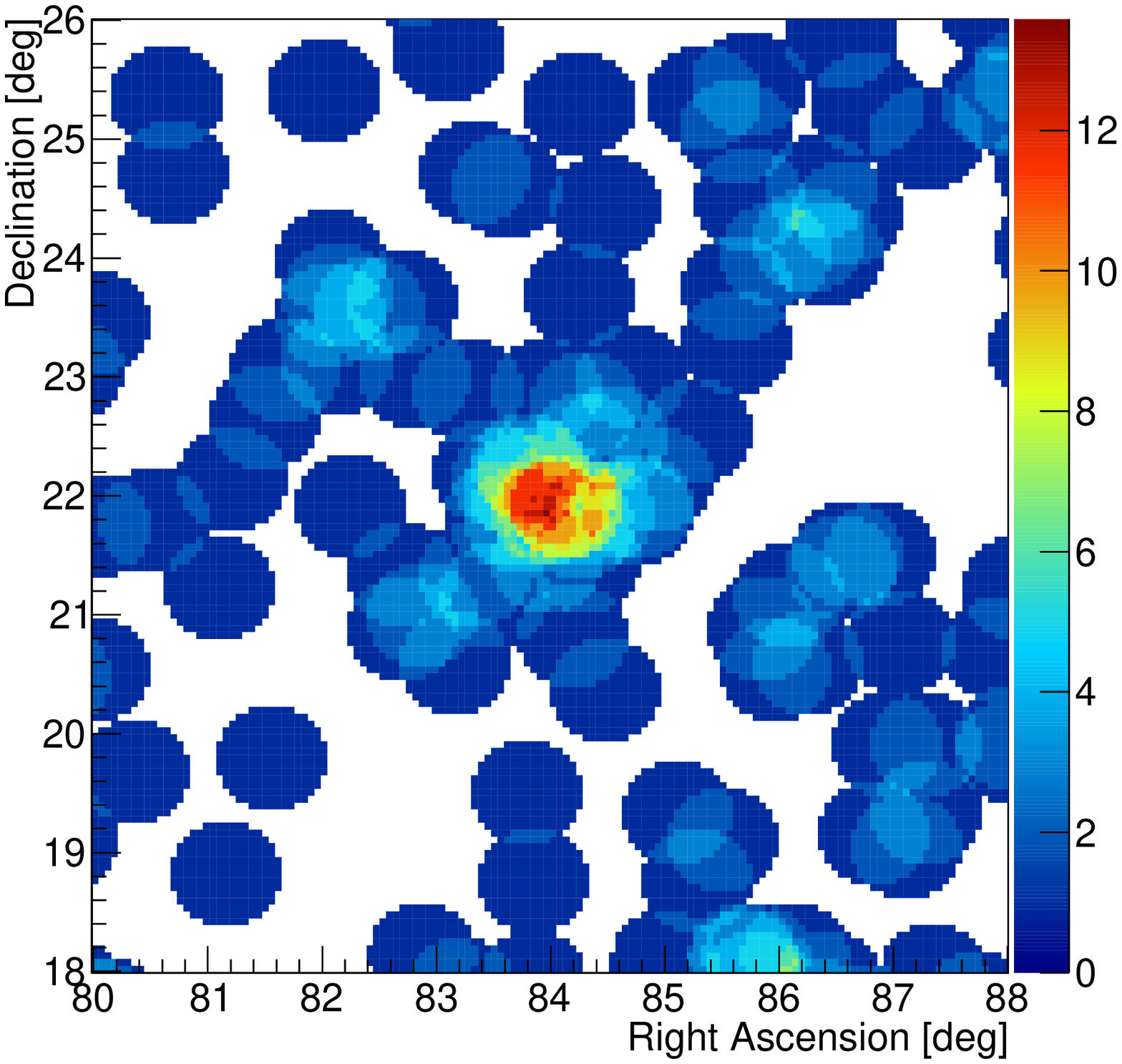}
\caption{Skymap in equatorial coordinates in the vicinity of the Crab Nebula
in 105 days of data with HAWC-250 (shown 
in the J2014/15 epoch). 
Events with more than 85\% of the available PMTs have been used.
The color scale shows the number of events detected within a 
0.45$^\circ$ circle around each point in the sky.
The left figure shows the region with no cuts and is completely
dominated by hadronic cosmic rays. The slight 
drift in rate is due to the changing acceptance of the 
detector, located at +19$^\circ$N latitude, for events further
from zenith. The figure on the right 
shows the same skymap after a strong cut which removes all but 1 in $\sim$10$^4$
background events. 
After cuts, at the location of the Crab,
we observe 12 events with an expected background of 0.81$\pm$ 0.05 
in this sample.}
\label{crabfield}
\end{figure}

The efficacy of the hadron rejection is shown most convincingly with 
real data from the instrument. Figure \ref{crabfield} shows
the region around the Crab in 105 days of data with HAWC in its
250-WCD configuration (HAWC-250). 
For the treatment here, we look at the largest, highest-energy
events in HAWC, events which have at least 85\% of the available
PMTs seeing light. This cut restricts the energy of photon events
to roughly $>$10 TeV.
Data are shown with a 0.45$^\circ$ top hat smoothing, counting how many
events occur within a 0.45$^\circ$-radius circle around the points shown. 
The left
panel of Figure \ref{crabfield} shows data with more than
85\% of the available PMTs seeing light. The right panel shows the 
same region of the sky with a strong cut on the core fit goodness-of-fit. 
The background is suppressed by about 10$^{-4}$ with $\sim$25\% 
efficiency for photons. At these extreme cut levels 
(very large, very gamma-like events), events are very rare. We
record 12 events (in the 0.45$^\circ$ bin) from the direction of the
Crab Nebula in 105 days. 
Using events recorded more than 1$^\circ$ from the Crab, 
we estimate 0.81$\pm$0.05$_{\rm{stat}}$ background events.

\begin{figure}
\includegraphics[width=0.45\linewidth]{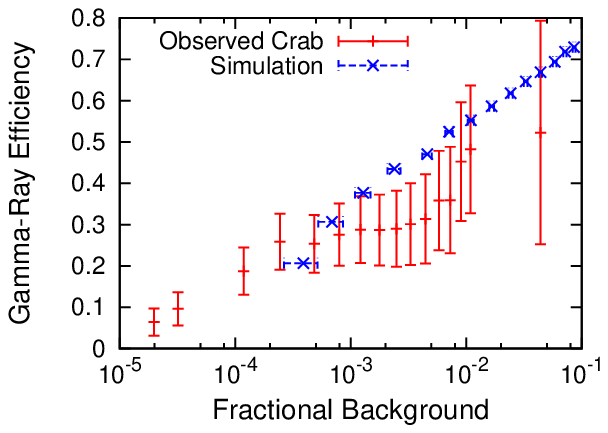}
\includegraphics[width=0.45\linewidth]{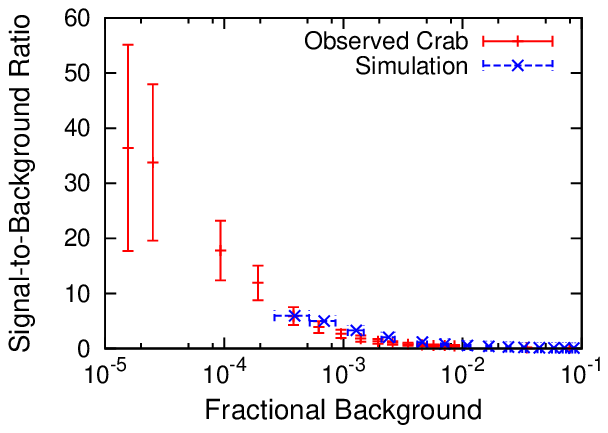}
\caption{These figures exhibit the performance of the photon/hadron
discrimination
described in the text. The left panel shows the fraction of the
gamma-ray signal expected and measured 
as a function of the fraction of the background
remaining as a cut on the core reconstruction goodness-of-fit is 
changed. Reasonable cuts produce a background acceptance of 10$^{-4}$ while
maintaining an efficiency for gamma-rays of $\sim$25\%. The right 
panel shows the signal-to-background ratio achieved as a function
of the fraction of background remaining after cuts.}
\label{cuttingonthecrab}
\end{figure}

Figure \ref{cuttingonthecrab} shows how the Crab excess depends on 
background rejection. We show measured and predicted
event passing rates as a function of the remaining background. 
The shape agrees well between data and simulation. Because it is time consuming
to simulate a full 100 days of experimental data, the
background estimate becomes statistically limited at a weaker
cut than data. Given the agreement at less-tight
levels, we are confident that $\sim$25\% gamma-ray efficiency
with only $10^{-4}$ background contamination is achievable, even at this early 
stage in the analysis. Stronger rejection, and a more pure photon sample, 
appears possible with harder cuts. 

\section{Isotropic Gamma-Ray Limit}
\label{thelimit}

\begin{figure}
\includegraphics[width=0.95\linewidth]{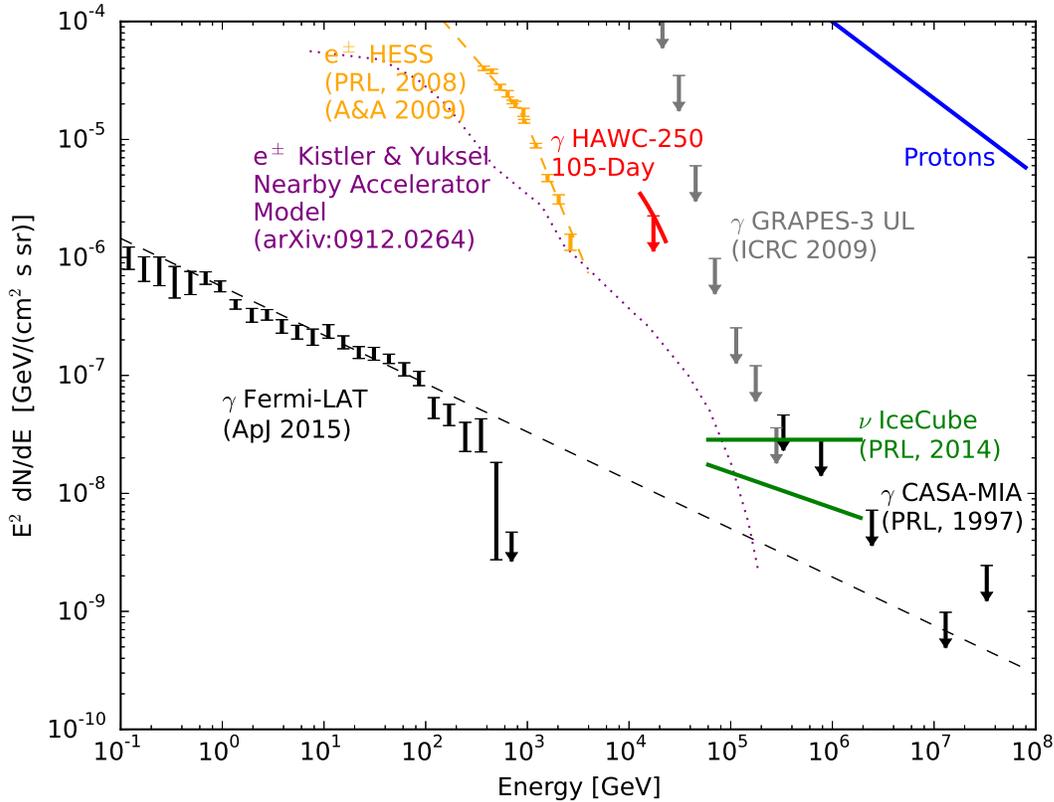}
\caption{Isotropic diffuse particle limits and measurements of a variety of
species 
\cite{icecubenu2014b}
\cite{hesscrelectrons} 
\cite{fermiisotropic2015}
\cite{casamia1997}
\cite{grapesicrc2009}
and an electron/positron model from 
\cite{pulsarelectrons2004}. 
Isotropic gamma-rays have been identified up to $\sim$ 1 TeV by 
the Fermi-LAT instrument. Above 1 TeV, the most constraining limits come 
from the GRAPES-3 instrument (in the 10-100 TeV range) and the CASA-MIA 
experiment (above 100 TeV). The early 
HAWC limits (shown in red) surpass the GRAPES-3 limits
in the 11-23 TeV energy. 
}
\label{isotropiclimit}
\end{figure}

The strength of the photon/hadron rejection allows us 
to set a simple, robust limit on an isotropic diffuse gamma-ray
flux.  Under no circumstances can an isotropic gamma-ray population
overproduce the observed off-Crab events. Furthermore, we have a bright
gamma-ray source of known flux to normalize our observations. 

Consider $F_{\rm{Crab}}$, the point-source 
differential flux of the Crab at some energy. Assume we measure
$S$ events from the Crab in some sample of time.
Let us assume the 
isotropic diffuse emission has the same spectral shape and is given by
$\Phi_{\rm{iso}}$ at the same energy as we measure the Crab.
Since the isotropic flux is normalized per unit solid angle, 
a small patch of sky with solid angle 
$\Omega_{\rm{bin}}$ will produce as many photons as a point source
with a flux $\Phi_{\rm{iso}}\Omega_{\rm{bin}}$. If, in the 
experiment, we record no more than $B$ events in an $\Omega_{\rm{bin}}$ sized
bin, and as long as we are looking near enough to the Crab that the instrument
response is flat, 
then we can say that the isotropic population of photons can be no 
higher than:

\begin{equation}
\Phi_{\rm{iso}} \leq {F_{\rm{Crab}}  \over \Omega_{\rm{bin}}} {B \over S}
\end{equation}

From the Crab observation in Figure \ref{crabfield}, we recorded
12 total events at the Crab location, with a background of
0.81$\pm$0.05 in a $0.45^\circ$ round bin. 
There will be some photons from the Crab outside our $0.45^\circ$ bin. 
The bin size was chosen to maximize the statistical significance of the 
Crab detection and, for a Gaussian-distributed point spread function, that
optimum includes only about 70\% of the photons. Given the systematic 
uncertainties
introduced by trying to account for these lost events, at this early stage 
in the analysis, we conservatively perform the limit calculation
with the assumption that the total number of Crab events is 11.2.

At 90\% confidence, we expect the true mean signal to background ratio
on the Crab to be larger than $S/B$=8.4. Using the H.E.S.S. fit of the
differential flux of the Crab \cite{hesscrab}, we obtain the limit
8.0$\times 10^{-15}~{\rm{cm^{-2}}}~{\rm{s^{-1}}}~{\rm{GeV^{-1}}}~{\rm{sr}}^{-1}$ at 17 TeV.
Using the HAWC-measured Crab
flux yields a similar limit. As we are still evaluating the HAWC
systematic errors, we choose to anchor the analysis on a published Crab flux
for this study. 

The HAWC limit is 
shown in Figure \ref{isotropiclimit}
along with 
a number of isotropic
measurements, limits and models for photons and other particle species.
We quote a HAWC limit from $\sim$11-23 TeV, the simulated energy range 
expected from the Crab for the cuts in this analysis. 

\section{Discussion and Prospects}
\label{discussionandprospects}

The limit presented here is very conservative and essentially comes
``for free'' from the HAWC observation of the Crab Nebula. It is not
representative of what is possible with HAWC, and a number of improvements
are possible:

\begin{itemize}

\item{HAWC data have the potential to constrain 
higher-energy photons, but we restrict the range of the claimed
limit to the energy range over which the
Crab is detected in this analysis.
As the modeling of the instrument is better understood,
this limit will be extended to lower and higher energies.
The sensitivity of HAWC at 100 TeV should be at least an 
order of magnitude
better than at 10 TeV. }

\item{We make no attempt to ascribe the observed off-Crab events to 
hadronic cosmic-rays. This assumption is robust, but very conservative.
A stronger limit is possible if
some fraction of the remaining background can be demonstrated, through
simulation, to be cosmic-ray background. As we understand HAWC
better, this will be possible and will improve the sensitivity substantially. 
Note that we must have some understanding of the background {\it{a priori}}
in order to make a positive detection. The analysis presented here
can only yield an upper limit.}

\item{We choose a photon/hadron separation cut 
that results in a convincing Crab detection
with an excellent signal to background ratio. As we acquire more data, we
will be able to cut harder and further enhance the purity of our sample. 
Figure \ref{cuttingonthecrab} suggests that 
more pure gamma-ray samples are possible. This stands to further 
improve the results.}

\end{itemize}

The first HAWC limit improves upon the current best limit in 
the 11-23 TeV range by the GRAPES-3 instrument.
Ultimately, we expect to achieve a limit below the level of the IceCube
diffuse neutrino flux at 10-100 TeV. 
HAWC is able to distinguish muons in an air shower
across essentially all of its 20,000 m$^2$ area, more than a factor of 10
larger than CASA-MIA or GRAPES-3. 

The 
limit presented here 
is formed using the background around the Crab Nebula. Strictly speaking,
this limit does not apply to any area of the sky except the immediate vicinity
of the Crab. Extending this analysis to the rest of the sky is underway.

Nevertheless, the present limit is the strongest in its energy range. With all 
the improvements discussed here, future observations will provide a new 
unique 
constraint
on the combined electron/positron/photon flux.
Above 100 TeV, our observations may be able, in the near future, 
to constrain the 
origin
of the IceCube astrophysical neutrino population. 

\section*{Acknowledgments}
\footnotesize{
We acknowledge the support from: the US National Science Foundation (NSF);
the US Department of Energy Office of High-Energy Physics;
the Laboratory Directed Research and Development (LDRD) program of
Los Alamos National Laboratory; Consejo Nacional de Ciencia y Tecnolog\'{\i}a (CONACyT),
Mexico (grants 260378, 55155, 105666, 122331, 132197, 167281, 167733);
Red de F\'{\i}sica de Altas Energ\'{\i}as, Mexico;
DGAPA-UNAM (grants IG100414-3, IN108713,  IN121309, IN115409, IN111315);
VIEP-BUAP (grant 161-EXC-2011);
the University of Wisconsin Alumni Research Foundation;
the Institute of Geophysics, Planetary Physics, and Signatures at Los Alamos National Laboratory;
the Luc Binette Foundation UNAM Postdoctoral Fellowship program.
}

%% \begin{bibliography}{99}

%% \end{bibliography}

\bibliographystyle{JHEP}
\bibstyle{JHEP}

\bibliography{bibliography}

\end{document}